\documentclass[twocolumn]{aastex7}

\usepackage{float}
\usepackage{xcolor}
\usepackage{caption}
\usepackage{subcaption}
\usepackage{graphicx}
\usepackage{amsmath}
\usepackage{multirow}
\usepackage[toc,page]{appendix}
\usepackage{natbib}
\usepackage{booktabs}
\usepackage{hyperref}

\begin{document}

\title{First Observation of CO$_2$ Foreground Absorption Toward the Galactic Center with JWST}

\author[0009-0006-5530-8354]{Jiawei Qiu}
\email{jqiu@astro.ucla.edu}
\affiliation{University of California, Los Angeles, Los Angeles, Department of Physics and Astronomy, CA, 90095-1547, USA}

\author[0000-0001-5800-3093]{Anna Ciurlo}
\email{ciurlo@astro.ucla.edu}
\affiliation{University of California, Los Angeles, Los Angeles, Department of Physics and Astronomy, CA, 90095-1547, USA}

\author[0000-0002-6753-2066]{Mark R. Morris}
\email{morris@astro.ucla.edu}
\affiliation{University of California, Los Angeles, Los Angeles, Department of Physics and Astronomy, CA, 90095-1547, USA}

\author[0000-0002-6126-3264]{Pierre Vermot}
\email{}
\affiliation{LIRA, Observatoire de Paris, PSL Research University, CNRS, Sorbonne Université, Université Paris Cité, 5 place Jules Janssen, 92195 Meudon, France}

\author[0000-0003-3920-8063]{Jacques Le Bourlot}
\email{}
\affiliation{LUX, Observatoire de Paris, PSL Research University, CNRS, Sorbonne Universités, 75014 Paris, France.}
\affiliation{Université Paris-Cité}

\author[0000-0002-2352-1736]{Daniel Rouan}
\email{}
\affiliation{LIRA, Observatoire de Paris, PSL Research University, CNRS, Sorbonne Université, Université Paris Cité, 5 place Jules Janssen, 92195 Meudon, France}



\author[0009-0003-7749-1864]{Aditya Togi}
\email{}
\affiliation{Department of Physics, 601 University Dr., Texas State University, San Marcos, TX 78666, USA}

\author[0000-0001-9554-6062]{Tuan Do}
\email{}
\affiliation{University of California, Los Angeles, Los Angeles, Department of Physics and Astronomy, CA, 90095-1547, USA}

\author[0000-0003-3230-5055]{Andrea M. Ghez}
\email{}
\affiliation{University of California, Los Angeles, Los Angeles, Department of Physics and Astronomy, CA, 90095-1547, USA}

\author[0000-0003-1532-7818]{Emeric Bron}
\email{}
\affiliation{LUX, Observatoire de Paris, PSL Research University, CNRS, Sorbonne Universités, 75014 Paris, France.}

\author[0000-0001-8738-6724]{Franck Le Petit}
\email{}
\affiliation{LUX, Observatoire de Paris, PSL Research University, CNRS, Sorbonne Universités, 75014 Paris, France.}

\author[0000-0002-8382-2020]{Yann Clénet}
\email{}
\affiliation{LIRA, Observatoire de Paris, PSL Research University, CNRS, Sorbonne Université, Université Paris Cité, 5 place Jules Janssen, 92195 Meudon, France}

\author[0000-0001-8782-1992]{Elisabeth A.C. Mills}
\email{}
\affiliation{Department of Physics and Astronomy, University of Kansas, 1251 Wescoe Hall Drive, Lawrence, KS 66045, USA}

\author[0000-0001-9611-0009]{Jessica R. Lu}
\email{}
\affiliation{Department of Astronomy, University of California, Berkeley, USA}

\begin{abstract}
CO$_2$ is an important, stable, and abundant molecule in the Universe, but it is very difficult to detect because it has no observable pure rotational transitions.
The unique sensitivity and resolution of the James Webb Space Telescope (JWST) provide a fresh way to investigate it. 
CO$_2$ is typically found in the solid phase (ice) on grain mantles in dense molecular clouds, but is less commonly detected in the gas phase (compared to common molecules such as CO and H$_2$O) and has mostly been found in protostellar and proto-planetary environments. 
Here, we report and characterize the first observations of gas-phase CO$_2$ absorption toward two IR-bright regions of the Galactic Center, thanks to the high sensitivity of JWST.
Using an LTE model we find a CO$_2$ gas excitation temperature between 20 and 50~K, a  column density around 2$\times$10$^{15}$~cm$^{-2}$ and a radial velocity consistent with 0. 
 We also report: 1) simultaneous detections of C$_2$H$_2$ and HCN absorption bands (near 13.7 and 14.0 $\mu$m, respectively), with column densitiy ratios of 1:3 and 3:2 with respect to gas-phase CO$_2$, and 2) CO$_2$ ice absorption with a ice-to-gas ratio of 90, consistent with previous findings. 
 We conclude that the absorbing medium is likely in the foreground, most likely from one or more somewhat clumpy cloud(s), located  between 0.15 and 4~kpc away from Earth.
 Additionally, we detected point-like CO$_2$ emission likely associated with a Galactic Center star (IRS~11SW), which is also spatially coincident with a previously reported X-ray source, raising the possibility that the system is a symbiotic binary.
\end{abstract}


\section{Introduction} \label{sec:introdiction}
Interstellar CO$_2$ plays an important role in the chemistry of the interstellar medium (ISM) as an abundant and stable molecule.
The importance of CO$_2$ extends beyond its abundance: it is a key player in the network of reactions that regulate the chemical abundances of molecules, and can help probe gas-grain interactions (e.g. \citealt{vanDishoeck+96}). 
However, CO$_2$ detection has long faced challenges. 
Unlike many interstellar species, it lacks a permanent dipole moment, meaning its rotational transitions are forbidden and thus hard to detect. 
Its detection has relied on vibrational bands in the mid-infrared, most notably through the 14.98~$\mu$m $\nu_2=0\rightarrow1$ Q-band line cluster and associated P- and R-branch features.
This introduces significant challenges, including difficulties in isolating the signal from strong dust continuum emission and disentangling it from overlapping spectral features \citep{vanDishoeck+96}.
Moreover, while warm and excited regions (such as protostellar disks and shocks) can enhance CO$_2$ emission, the varying excitation conditions and optical depth effects complicate the quantitative interpretation of such line measurements \citep{Boonman2003}.

In molecular clouds, most CO$_2$ is expected to be in ice form (e.g. \citealt{vanDishoeck+96}, and references therein).
Gas-phase CO$_2$ can be observed both in emission and absorption, and the physical conditions giving rise to either differ significantly.
Emission is typically observed in regions with high temperatures and densities where CO$_2$ molecules are actively radiating.
For example, \citet{Boonman2003} reported observations of gas-phase CO$_2$ toward massive protostars, where the warm, excited gas produces infrared emission features. 
\cite{Boonman+03b} reported on CO$_2$ emission toward Orion-KL, produced by shock-heated gas in protostellar environments. 
In contrast, absorption features arise when a cooler, foreground gas lies along the line of sight to a bright background. 
For example, \citet{An_2009} detected CO$_2$, C$_2$H$_2$, and HCN absorption in young stellar object candidates in the galactic center, revealing the presence of these molecules in the circumstellar medium.
These examples illustrate the diverse physical conditions under which CO$_2$ can be detected, providing insight into both warm, dynamic regions and quiescent interstellar gas.

In this work, we leverage the enhanced sensitivity and spectral resolution of JWST to investigate CO$_2$.
We report the first detection of gas-phase CO$_2$ absorption toward the Galactic Center.
The gas-phase CO$_2$ absorption is spatially extended and we also detect absorption from HCN, C$_2$H$_2$, and CO$_2$ ice.
The observed velocities indicate that the absorbing gas is most likely located in the near foreground, within a few kpc of Earth. 
Additionally we also detect a point-like CO$_2$ emission source, mostly likely associated with a star near the Galactic Center.

The paper is organized as follows: Section~\ref{obs} describes the two fields we observed using MIRI/MRS, Section~\ref{sec:analysis} describes the methods and models we used to determine the physical properties of the molecular species, Section~\ref{sec:results} presents the results, and Section~\ref{sec:discussion} discusses the implications of our findings. Finally, our conclusions are summarized in Section \ref{sec:conclusions}.

\section{Observations} \label{obs}
 
\begin{figure}[tb]
    \centering
    \includegraphics[width=\linewidth]{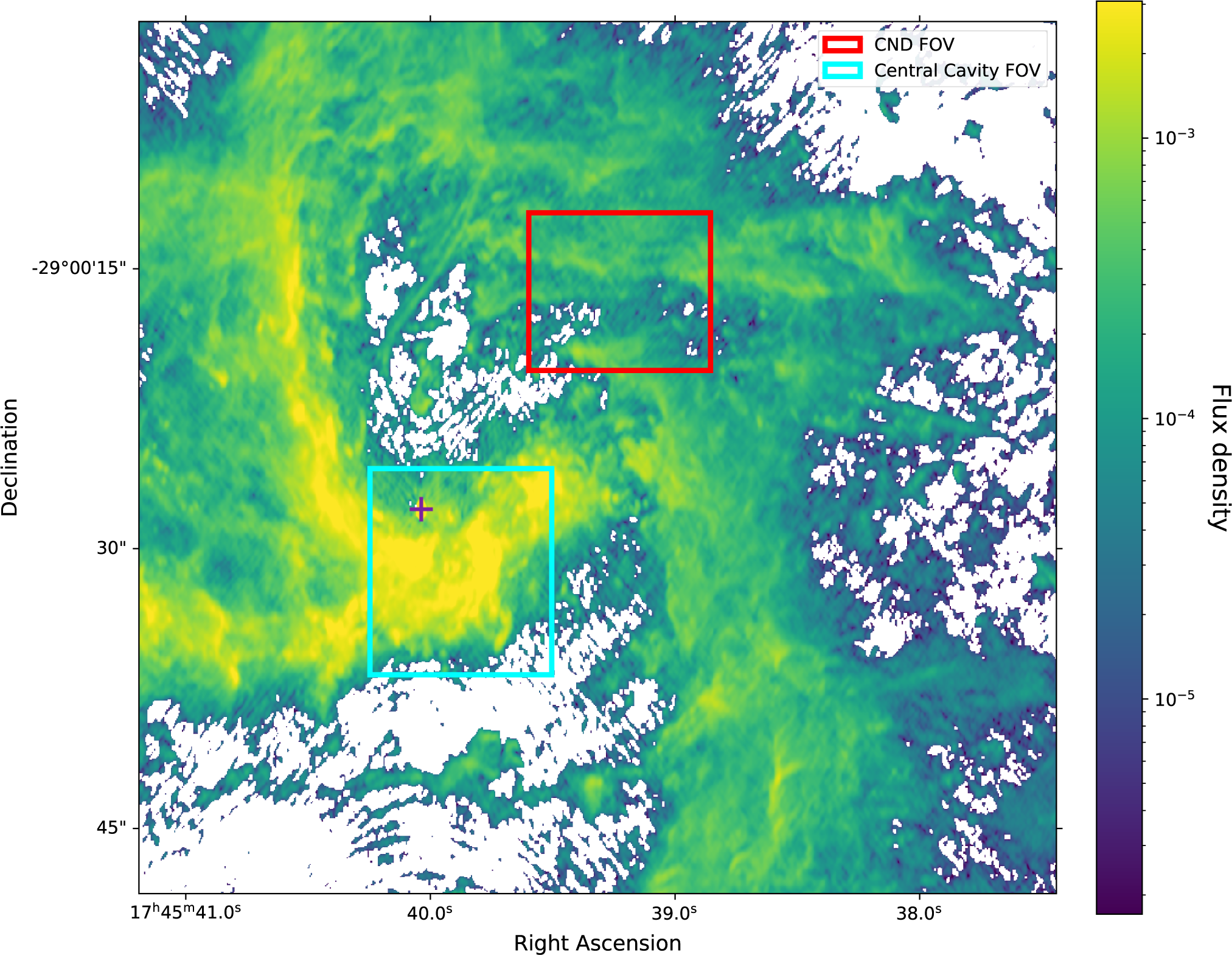}
    \caption{The observation fields overlaid upon a JVLA 2cm continuum map of the Galactic Center, from \citet{Morris+17}. The CC and CND pointings are represented by the cyan and red rectangles, respectively. The location of Sgr A* is marked by the magenta cross.}
    \label{fig:observation_fields}
\end{figure}

We observed two fields with JWST's mid-infrared integral field spectrograph MIRI/MRS in Cycle 2 (GO program number 3166, PI: A. Ciurlo).
Our observational program focused on two fields, one in the Circumnuclear Disk (CND) and the other on the central cavity (CC) of the CND, as shown superimposed on a 2~cm radio continuum map (from \citealt{Morris+17}) in Figure~\ref{fig:observation_fields}.
The observations took place on April 18$^{\text{th}}$ and September 11$^{\text{th}}$, 2024 UTC, respectively. 
The first field is located 18.25$''$ northwest of Sgr~A* (centered on R.A. $17^\text{h}45^\text{m}39\rlap{.}^{\text{s}}12$, Decl. $-29^{\circ}00'16\rlap{.}''26$ J2000, hereafter referred to as the CND field).
The second field is centered 3.29$''$ south of Sgr~A* (centered on R.A. $17^\text{h}45^\text{m}39\rlap{.}^{\text{s}}97$, Decl. $-29^{\circ}00'31\rlap{.}''24$ J2000, hereafter referred to as the CC field).
Reference sky background fields were observed on September 16$^{\text{th}}$, 2023 and April 18$^{\text{th}}$, 2024 (centered on R.A. $17^\text{h}45^\text{m}10\rlap{.}^{\text{s}}27$, Decl $-29^{\circ}02'13\rlap{.}''67$ J2000) and on August 20$^{\text{th}}$, 2024, and September 11$^{\text{th}}$, 2024 (centered on R.A. $17^\text{h}44^\text{m}46\rlap{.}^{\text{s}}65$, Decl $-28^{\circ}53'44\rlap{.}''41$ J2000).
All fields were observed in all 4 MIRI wavelength channels.
For this analysis, we focus on channel 3 (11.55 to 17.98~$\mu$m), particularly the wavelength range near 15~$\mu$m, where the instrumental wavelength sampling interval is 0.0025~$\mu$m (50~km~s$^{-1}$), and the spatial pixel scale is 0.2~arcsec (R=$\lambda$/FWHM$\simeq$2700, PSF FWHM$\simeq$0.6$''$). The data described here may be obtained from the MAST archive (doi:\dataset[10.17909/h7sh-wn90]{https://dx.doi.org/10.17909/h7sh-wn90}).

The data were reduced through the standard JWST MIRI pipeline (\citealt{bushouse_2025_15178003}, versions 1.15.1 and 1.13.3 for the CC and CND exposures, respectively) with the additional application of \texttt{fit\_residual\_fringes\_1d}\footnote{For more detailed description on fringe removal, see \url{https://jwst-docs.stsci.edu/known-issues-with-jwst-data/miri-known-issues/miri-mrs-known-issues}.} to spectra of every pixel to remove the instrumental fringing effect (as described in \citealt{Vermot+25}, a known issue affecting MIRI data).
The JWST calibration pipeline includes processes such as detector-level corrections, flat-fielding, spectrophotometric calibration, mosaicking, and resampling to produce science-ready data cubes.
We also note that although there was a complementary sky observation, sky background subtraction is not applied to the data analyzed in this work, as background subtraction is not as good due to the rapidly varying background emission in the region.

\begin{figure*}[tb]
    \centering
    \includegraphics[width=0.8\textwidth]{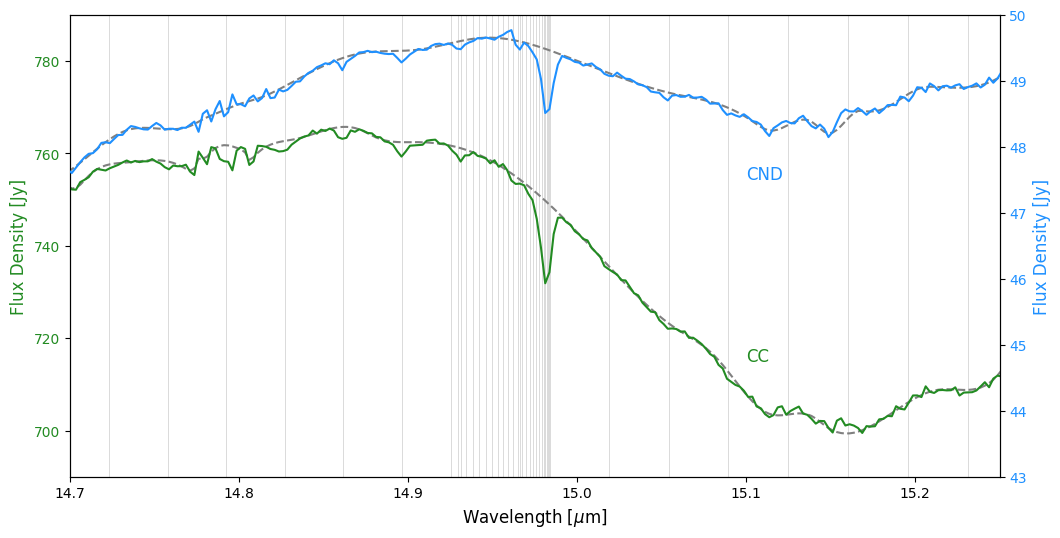}
    \caption{Integrated spectra over each of the two fields showing the detection of the CO$_2$ absorption feature. The spectrum extracted from the Circumnuclear Disk (CND) field is shown in blue and that from the Central Cavity (CC) pointing is shown in green, with their corresponding scales on the right and the left, respectively. The dark gray, dashed lines show the baseline model (see Section~\ref{sec:analysis} for details). The vertical light gray lines indicate the wavelengths of modeled transitions. The acetylene and hydrogen cyanide features are at shorter wavelengths (not shown -- see Figure~\ref{fig:hcn_and_acetylene}), while part of the broad solid CO$_2$ feature can be seen to the right of this Figure~(see Figure~\ref{fig:co2_ice_spectrum}).}
    \label{fig:integrated_spectrum}
\end{figure*}

We extracted a spectrum for each pointing by integrating over the entire field of view ($\sim11''\times12''$ for the CC exposure, and $\sim11''\times9''$ for the CND exposure), masking out the edge pixels (two pixels from the edge), which contain obvious artifacts.
The extracted spectra of both fields are shown in Figure~\ref{fig:integrated_spectrum}.
Both fields show prominent CO$_2$ absorption features, the most pronounced being the central Q branch (with a signal-to-noise ratio $\sim$30).
Additionally, a few isolated absorption lines from the P and R branch transitions can be seen around the central feature.\\

\section{Analysis} \label{sec:analysis}
\subsection{Absorption modeling}

In order to describe the properties of the gas that produces the absorption features, we created a model assuming local thermodynamic equilibrium (LTE) populations among the rotational states of the ground vibrational state.
See Appendix\,\ref{NLTE_Model} for a discussion of possible non-local thermodynamic equilibrium effects.
The model has four parameters that can freely vary: radial velocity with respect to the local standard of rest (LSR), intensity normalization, rotational temperature, and intrinsic velocity dispersion.
The model includes CO$_2$ ro-vibrational transitions originating from rotational states with $0\leqslant J\leqslant 50$ in the $\nu_2=0$ vibrational state to higher energy states with $\nu_2=1$ and $\Delta J=0,\pm1$. 
The choice of $J=50$ for the highest rotational state considered is justified by the fact that higher rotational states would only be significantly populated at temperatures ($>$ 1000~K) unlikely to be present in the absorbing medium.
The flux density of the LTE model at a certain wavelength is calculated using the following equation, in which we assume intrinsically Gaussian profiles:
\begin{equation}
    \label{eq1}
    \begin{aligned}
    I(\lambda)=&I_0(\lambda)-\sum_i \bigg[KB_i(g_ie^{-\frac{E_i}{kT}})
    \\
    &\exp\left(-\frac{[\lambda-\lambda_i(1+v/c)]^2}{2(\lambda_i\delta v_i/c)^2}\right)\bigg]
    \end{aligned}
\end{equation}
where $I_0(\lambda)$ is the continuum value at wavelength $\lambda$, $K$ is the overall normalization related to the column density of the gas, $T$ is the rotational temperature of the gas, $c$ is the speed of light, $v$ is the LSR velocity, and $B_i, g_i, E_i, \lambda_i, \delta v_i$ are the Einstein B coefficient, lower level statistical weight, lower level energy, rest wavelength, and intrinsic velocity dispersion of the i$^{th}$ transition, respectively.
The transitions and constants associated with come from the HITRAN database\footnote{\url{https://hitran.org/data-index/}}.
Other molecular transitions and constants in this work were also acquired from the HITRAN database.
$\nu_2$=0-1 transitions were used in modeling of CO$_2$, $\nu_5$=0-1 for acetylene, and $\nu_2$=0-1 for hydrogen cyanide \citep{2022JQSRT.27707949G}.
For simplicity, all transitions are assumed to have an intrinsically Gaussian profile. 
$\delta v_i$ is the observed velocity dispersion which corresponds to:
\begin{equation}
    \label{eq2}
    \delta v_i=\sqrt{\delta v_{\text{intrinsic}}^2+\delta v_{\text{instrumental}}^2}
\end{equation}
where $\delta v_{\text{intrinsic}}$ is the intrinsic velocity dispersion, assumed constant for all transitions,  $\delta v_{\text{instrumental}}^2$ is the instrumental width at the wavelength of the i$^{th}$ transition. The wavelength-dependent instrumental width of MIRI/MRS is calculated using the empirical formula given by \citet{Argyriou2023}, $R(\lambda)=4603-128\times\lambda$.

In order to correctly model the continuum emission throughout the wavelength range of interest, we identified and masked regions that have known absorption features.
To ensure that we are using the correct Doppler shift and intrinsic line width for the masking of the CO$_2$ transitions, we determined a rough estimate of the continuum baseline by first masking only the central Q branch cluster of transitions.
Subsequently, we applied the LTE model to the central part of the spectrum (14.96-14.99 $\mu$m).
This rough fit confirms that the radial velocity is close to zero and the line width is dominated by instrumental resolution (intrinsic line width $\sim20$~km~s$^{-1}$ compared to the $\sim50$~km~s$^{-1}$ instrumental width).
We then placed the masking boundaries around each transition at $\sim3\sigma$ times the instrumental resolution.
With this masking applied, we determined a more precise baseline using cubic B-spline (see Figure~\ref{fig:integrated_spectrum}).
We then applied the LTE model to the entire baseline-subtracted spectrum (see Figure~\ref{fig:co2_gas_modeling}). 

\subsection{Column density determination}
We also calculated the column densities of CO$_2$ from the two exposures using the equivalent width of the Q branch absorption in the two spectra integrated over each of the two fields of view.
We find that the parameter fitting results are also robust to the somewhat arbitrary selection of baseline-fitting wavelength range.
Assuming that the CO$_2$ absorption is optically thin (which is very likely since the absorption is weak, even after correcting for instrumental resolution), we made the approximation that the intrinsic optical depth is $\tau\approx\Delta I/I_0$, where $\Delta I$ is the maximum depth of the line and $I_0$ is the baseline without absorption.
We calculated the column density of the gas using equivalent width because it is not affected by instrumental resolution, and with the above assumption the equivalent width of the single $i^{\text{th}}$ transition is
\begin{equation}
\label{eq3}
\begin{aligned}
    EW_i&=\sqrt{2\pi}\frac{\Delta I_i\sigma_i}{I_{0,i}}\\
    &=\left(\frac{\sqrt{2\pi}\lambda_i^2}{2\sqrt{2ln(2)}c}\right)\left(\frac{2\sqrt{2ln(2)}c\Delta I_i\sigma_i}{I_{0,i}\lambda_i^2}\right)\\
    &\approx\left(\frac{\sqrt{2\pi}\lambda_i^2}{2\sqrt{2ln(2)}c}\right)\tau_i\Delta\nu_i
\end{aligned}
\end{equation}
where $\sigma_i$ is the standard deviation of the Gaussian corresponding to the $i^{\text{th}}$ transition, and $\Delta\nu_i$ is the FWHM of the $i^{\text{th}}$ absorption profile in frequency space.
The equivalent width of the Q branch absorption cluster can then be calculated using
\begin{equation}
\label{eq4}
    EW=(\text{All transitions})\sum_{\text{14.96 $\mu$m}}^{\text{14.99 $\mu$m}}\Delta I\Delta\lambda/I_0
\end{equation}
where $\Delta\lambda$ is the wavelength sampling interval.
From there, a column density is calculated using the following equation:
\begin{equation}
\label{eq5}
    N_i=\frac{8\pi^{3/2}}{2\sqrt{ln2}}\frac{1}{\lambda_i^2A_{ul}}\frac{g_l}{g_u}\tau_i\Delta\nu_i=\frac{8\pi c}{\lambda_i^4A_{ul}}\frac{g_l}{g_u}EW_i
\end{equation}
where $N_i$ and $EW_i$ are the column density and equivalent width corresponding to the $i^{\text{th}}$ lower state, with $EW_i$ calculated from $EW$ assuming thermodynamic equilibrium (e.g. \citealt{1971ApJ...169L..39R}).
In the calculation, we assume that the population in the upper ro-vibrational states is negligible, as collisions can only populate those states significantly at temperatures much higher than those likely to occur in the absorbing medium.
These lower states have $J$ between 0 and 30 (the contributors to the Q branch feature). 

The LTE assumption is likely to be quite good for the homopolar molecules, CO$_2$ and C$_2$H$_2$, because the absence of a dipole moment makes their rotational transitions essentially forbidden, so that the level populations are determined almost entirely by collisions at the ambient gas temperature.
However, for HCN, collisional excitation competes with spontaneous radiative rotational transitions, so unless the collision rate is extremely high (particle densities $>10^5$ cm$^{-3}$), the rotational levels of HCN become subthermally populated.
The rotational temperatures reported in Table \ref{table:params} are consistent with this expectation; that of HCN is substantially lower than those of both CO$_2$ and C$_2$H$_2$.
The sizable difference between the rotational temperatures of CO$_2$ and C$_2$H$_2$ has no obvious explanation, although the uncertainties are large enough that consistency between them cannot be ruled out.

After assigning the contribution of each transition to the main absorption feature according to the best-fit temperature, the total column density of CO$_2$ is calculated as the sum of the column densities of individual rotational states. 
The column densities of individual levels also serve as a posterior check on the assumption that $J$ up to 30 is reasonable, as the column density contribution from $J=30$ is 4 orders of magnitude less than that of $J=2$.
Alternative methods for measuring column density lead to the same results (see Appendix~\ref{appendix:A2} for details).
The resulting LTE model, best-fit parameters, and column density are shown in Section~\ref{sec:results}).
The LTE model fitting used the \texttt{scipy.optimize} package and the trust region reflective method, with generous bounds allowed: 0.1~km~s$^{-1}$ $\leqslant\delta v_{\text{intrinsic}}\leqslant$ 100~km~s$^{-1}$, $-$100~km~s$^{-1}$ $\leqslant v \leqslant$ 100~km~s$^{-1}$, and 0 K $\leqslant T \leqslant$ 500 K.

\subsection{Column density spatial variation}
In order to investigate the spatial variation of the emission across each field of view, we also fit spectra extracted from small apertures.
An aperture diameter of 5~px (corresponding to 1~arcsecond) was chosen because it offers a reasonable trade-off between spatial resolution and signal-to-noise ratio (the SNR of any such aperture is usually $\gtrsim$10, while in typical cases a single pixel only has SNR$\sim$3).
Since the derived parameters do not vary strongly across the field with this choice of aperture, 
we have chosen to fit the LTE model to spectra extracted from apertures with 5~px diameters around each pixel.
The use of 5-px apertures also helps smooth features smaller than the instrumental point spread function at 15~$\mu$m ($\sim0.6''$, or 3~px), subject to inherent noise in the data.
The resulting parameter maps are presented in Section~\ref{sec:results}).
The analysis applied to CO$_2$ lines described in this section was also performed for the simultaneously observed $v_2~=~0\rightarrow1$ Q-branch absorption lines of C$_2$H$_2$ near 13.7 $\mu$m and the P, Q, and R-branch lines of HCN at wavelengths near 14.0 $\mu$m.

\section{Results} \label{sec:results}

\subsection{CO$_2$, C$_2$H$_2$, and HCN properties}
\begin{figure*}[htb]
  \centering
  \includegraphics[width=0.8\textwidth]{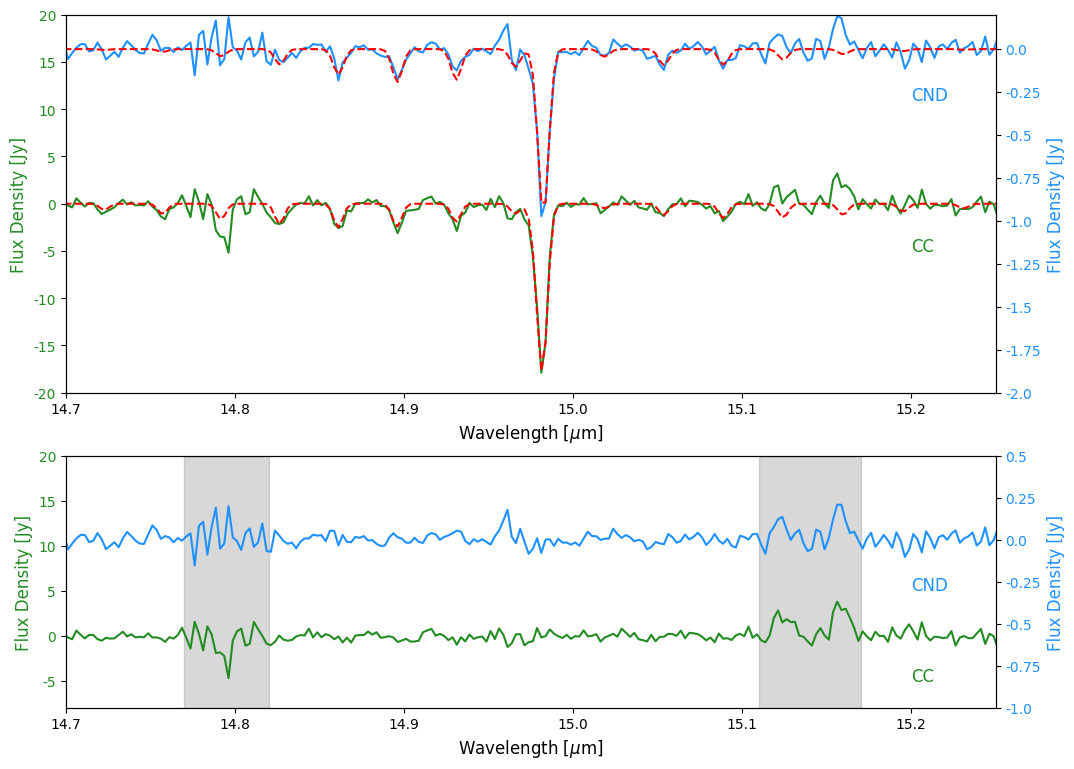}
  \caption{The top panel shows CO$_2$ absorption in the baseline-subtracted spectrum (solid line) along with the LTE model fit (dashed line), and the bottom panel shows model residuals for both pointings. Color and scale correspondences same as Figure~\ref{fig:integrated_spectrum}; the red dashed lines represent the best-fit LTE models. The two shaded columns in the bottom panel mark deviations from the model fit caused by the inherent noise in the data (close to 14.8 $\mu$m) and the inaccurate baseline tracing due to absorption from CO$_2$ ice (close to 15.15 $\mu$m).}
  \label{fig:co2_gas_modeling}
\end{figure*}

The LTE model and residual for the spectra extracted from each of the observed fields are plotted in Figure~\ref{fig:co2_gas_modeling}, and the best-fit parameters and derived column densities are reported in Table~\ref{table:params}.


We also performed a search for CO$_2$ using the sky background exposures. 
Inspecting the integrated spectra of the individual fields of view, we could detect no CO$_2$ absorption, and combining the four sky background exposures does not change that conclusion.
The sky background exposures also have a relatively higher ratio of RMS noise to the continuum level, at $\sim1\%$ compared to a noise ratio $\sim0.1\%$ for the science target exposures.
We therefore placed an upper limit on the equivalent width of the absorption in the background fields by assuming the maximal absorption intensity is equal to the standard deviation of the data in the wavelength range from 14.9 $\mu$m to 15.06 $\mu$m and that the gas temperature is roughly 30 K (to match that of the foreground gas observed in the science exposures).
The upper limit on the column density of CO$_2$ in the sky exposures is thus found to be  $\sim7\times10^{14}\text{ cm}^{-2}$, a factor of 3 lower than the column density reported for the science exposures.

\begin{figure*}[htb]
    \centering
    \includegraphics[width=0.8\textwidth]{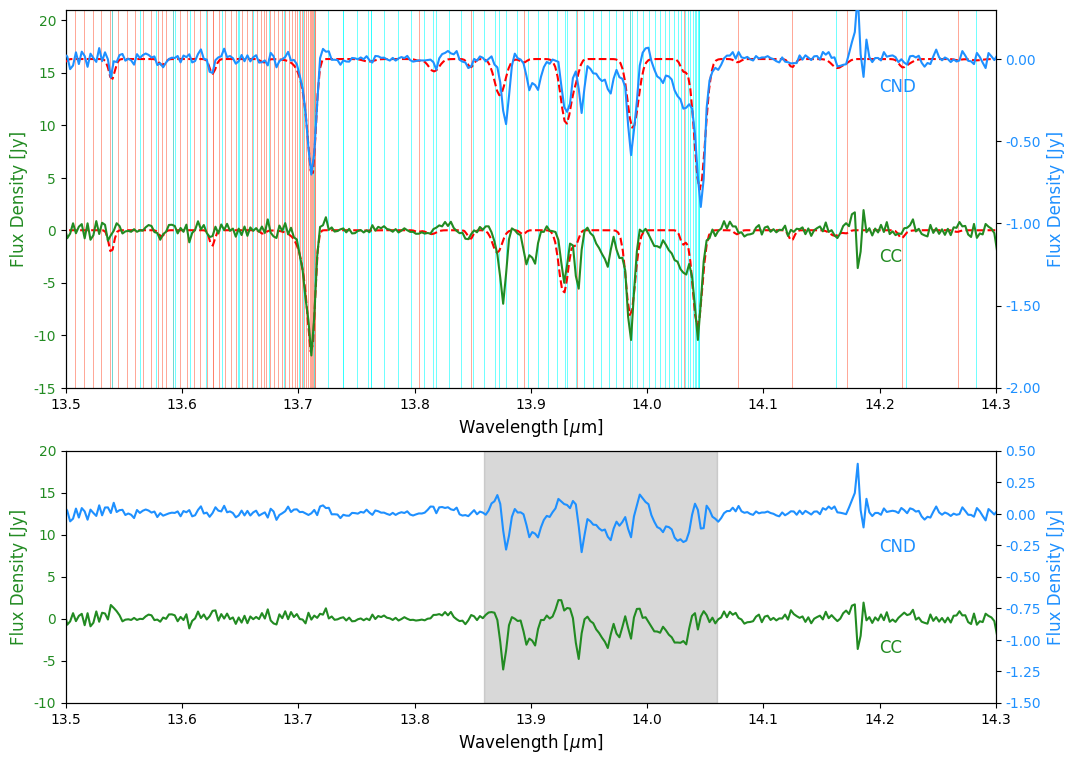}
    \caption{C$_2$H$_2$ and HCN absorption in the baseline-subtracted spectra, along with the LTE model fit (red dashed lines)(top panel), and model residuals for both pointings (bottom panel). The Q-branch transitions of C$_2$H$_2$ are around 13.7~$\mu$m, while those for HCN are around 14.05~$\mu$m. The orange vertical lines represent modeled C$_2$H$_2$ transitions, while the cyan vertical lines represent modeled HCN transitions. The shaded column in the bottom panel contains the region with absorption features not accounted for by the LTE model of the two molecules.}
    \label{fig:hcn_and_acetylene}
\end{figure*}
We also found C$_2$H$_2$ and HCN absorptions in the spectrum extracted from both exposures, as shown in Figure~\ref{fig:hcn_and_acetylene}. 
The absorption from these molecules is even weaker (signal-to-noise $\gtrsim$ 25) than that for CO$_2$, and the baseline modeling is made more arduous because the two species have many closely spaced absorption features. 
Due to their proximity, the two species were modeled simultaneously using the same local thermodynamic equilibrium assumption.
Using the parameters derived from the best-fit models, we calculated the column densities of these species and report them in Table~\ref{table:params} (note that the HCN column density uncertainties shown are only statistical, and given that there are unidentified features intermixed with the HCN transitions, these uncertainties are only lower limits).
The column densities reported for HCN suffer from greater uncertainty, as HCN has numerous transitions, making the baseline estimation more challenging.
It is clear from Figure~\ref{fig:hcn_and_acetylene} that there are features that the LTE model cannot account for between 13.85 and 14.05~$\mu$m, possibly caused by some other species that we have not identified or by an imperfect baseline fitting in the region with too many features.
We note that these features are unlikely to be caused by high-velocity HCN components, as models having two or even three velocity components hardly show any statistical improvements while requiring radial velocities exceeding 200~km~s$^{-1}$.


\begin{deluxetable}{c c c c c c}
\tablecaption{Best-fit parameters of the three molecular species \label{table:params}}
\tablehead{
  \colhead{Molecule} & \colhead{Pointing} & \colhead{Temperature} & 
  \colhead{Radial velocity} & \colhead{Velocity dispersion} & 
  \colhead{Column density} \\ 
  \colhead{} & \colhead{} & \colhead{(K)} & \colhead{(km s$^{-1}$)} & 
  \colhead{(km s$^{-1}$)} & \colhead{(10$^{15}$ cm$^{-2}$)}
}
\startdata
\multirow{2}{*}{CO$_2$} & CND & 22.8$^{+30}_{-12}$ & 0$\pm$8 & $\leqslant$55 & 1.60$\pm$0.3 \\
 & CC & 46.0$^{+39}_{-24}$ & $-$2$\pm$9 & $\leqslant$45 & 2.05$\pm$0.3 \\
\multirow{2}{*}{C$_2$H$_2$} & CND & 93$\pm$23 & +6$\pm$12 & $\leqslant$50 & 0.60$\pm$0.08 \\ 
 & CC & 105$\pm$32 & $-$4$\pm$11 & $\leqslant$40 & 0.57$\pm$0.08 \\
\multirow{2}{*}{HCN} & CND & 9.8$^{+15}_{-5}$ & +28$\pm$9 & $\leqslant$90 & 2.8$\pm$0.4 \\
 & CC & 6.5$^{+8}_{-3}$ & $-$19$\pm$9 & $\leqslant$90 & 1.8$\pm$0.3
\enddata
\end{deluxetable}

\subsection{CO$_2$ spatial distribution}
\begin{figure*}[htb]
    \centering
    \includegraphics[width=\textwidth]{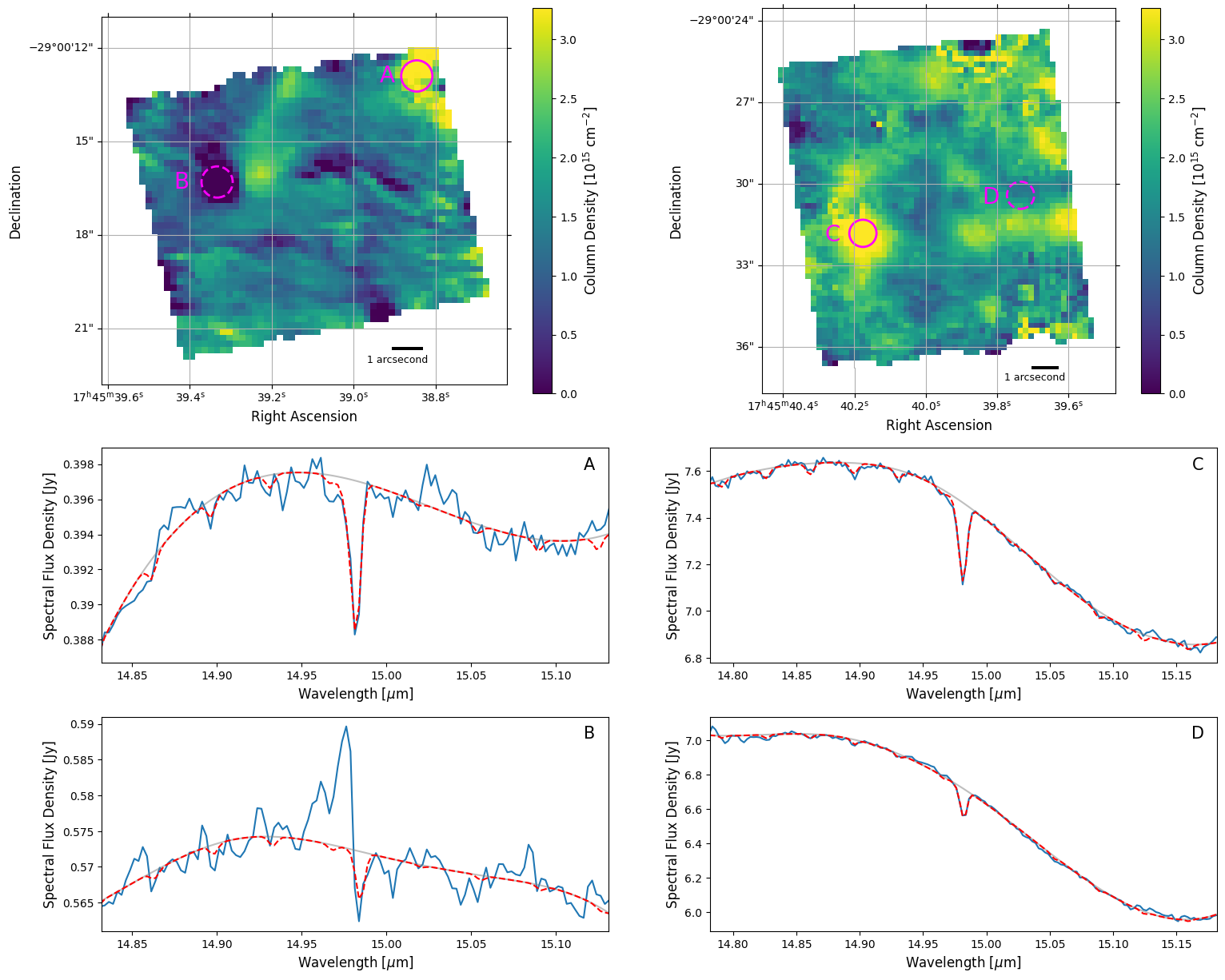}
    \caption{Column density maps and selected spectra of CO$_2$ absorption from the CND pointing (left column) and the CC pointing (right column). The upper spectra are from the solid magenta apertures and the lower spectra are from the dashed magenta apertures. Gray solid lines represent the baseline fits and the red dashed lines represent the best LTE model fit.}
    \label{fig:apertures}
\end{figure*}

Figure~\ref{fig:apertures} shows the maps of CO$_2$ column density derived using the main Q-band absorption feature resulting from the aforementioned spatial variation analysis.

\begin{figure*}[htb]
    \centering
    \includegraphics[width=0.8\textwidth]{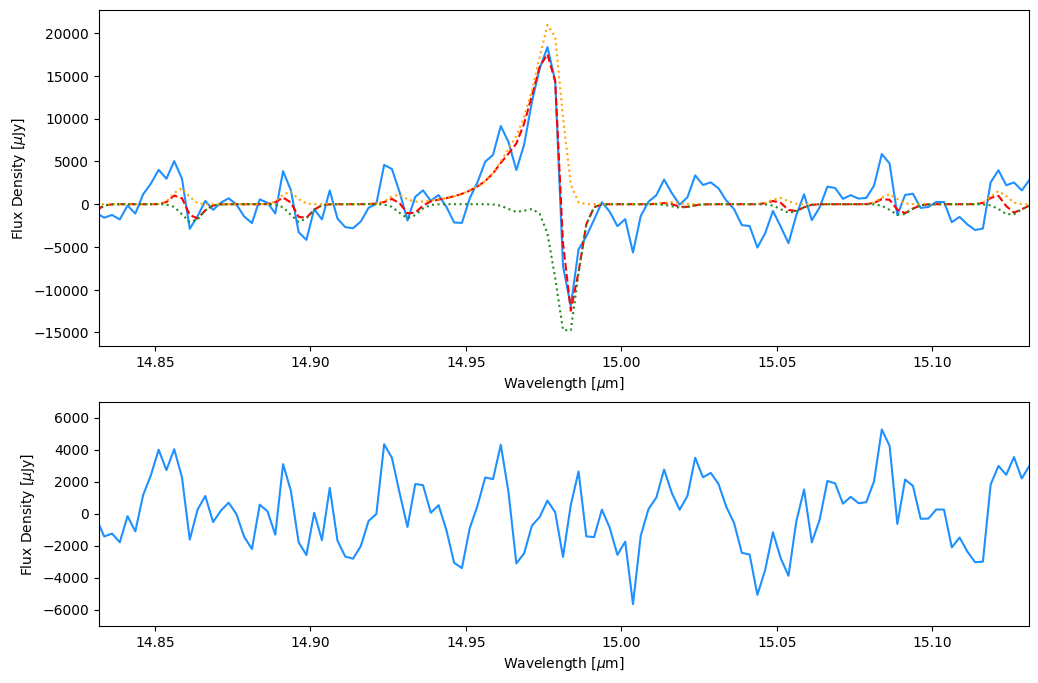}
    \caption{Top panel: baseline-subtracted emission spectrum (Aperture B, CND), overlaid with the emission model (orange dotted line), averaged absorption model (green dotted line), and the overall fit (red dashed line). Bottom panel: residual of the model.}
    \label{fig:emission}
\end{figure*}

With a few exceptions, these maps show extended absorption with variations up to a factor of 2.
Relatively lower signal-to-noise ratios at the edge of the maps could cause some edge effects, but that does not impact the overall picture of extended, relatively slowly varying absorption across the fields of view.
A noteworthy exception is an apparent ``dip'' in the column density map of the CND pointing. 
We identified {\em emission} from CO$_2$ as the cause of this feature in the equivalent width map.
The spectrum extracted from an aperture centered on this region, along with those extracted from apertures centered on the peak column density regions and an ``average'' region, are shown in Figure~\ref{fig:apertures} (the circles in the upper panels represent the apertures from which the spectra were extracted).
Indeed, we detect an emission feature corresponding to the aforementioned ``dip'' and that emission is very likely CO$_2$, given its peak wavelength (immediately next to the Q branch absorption features) and the expected low intensity of high-$n$ atomic hydrogen transitions 
close to the wavelength of interest.
To further investigate the nature of the emission, we used the average spectrum of four apertures equally spaced around the emission source at distances of $\sim$1.2 arcseconds as an estimate of the absorption profile at this location.
We then subtracted that absorption profile from the spectrum of the emission source to produce a spectrum free of absorption. 
After that, we applied an LTE emission model constructed in a fashion similar to that of the absorption model to the emitting component.
The resulting model fit to the emission, combined with the above-mentioned average absorption, is shown in Figure~\ref{fig:emission}.
We tried a few other methods to determine the radial velocity of the emission component, and depending on the distance of apertures placed around the emission source to get an average absorption model, baseline determination, and if the absorption component is allowed to freely vary, these methods yielded radial velocities consistent with each other within 15~km~s$^{-1}$, and we use that as an estimate of the uncertainty in the inferred velocity.
We conclude that the emission arises from a point-like source that has a LSR velocity of $-70\pm15$~km~s$^{-1}$.

For completeness, we conducted a search in both fields for other possible CO$_2$ emission sources, integrating the flux at all locations between 14.93 and 15.0 $\mu$m that are more than 3-$\sigma$ higher than the deduced baseline at the location.
This search concluded that in the two fields, there is no significant CO$_2$ emission at any location other than the aforementioned point-like emission source.
There is a small residual peak in the CND spectrum shown in Figure~\ref{fig:co2_gas_modeling} blue-ward of the main absorption feature, but it does not specifically appear at any one location, and if it were a real CO$_2$ feature, it would have a velocity of $\sim-450$~km~s$^{-1}$, which does not correspond to the velocity of any known atomic, ionic, or molecular feature.
On the other hand, the HI n=16-10 recombination line has a rest wavelength exactly at the location of the residual peak (14.962~$\mu$m).
To assess whether the intensity of this feature is consistent with this identification, we compared its flux to that of the HI 7-6 recombination line at 12.372~$\mu$m, finding a flux ratio of $\sim50$.
This ratio is in good agreement with the calculated flux ratio between these lines for a wide range of assumed temperatures from $10^3$ to $10^{4.3}$~K and for densities in the range $10^{3-7}$ cm$^{-3}$, based on the emissivity tabulations of \citet{1995MNRAS.272...41S} and  \citet{2015A&A...573A..42L}.
These ranges encompass the values found in the ionized gas of the CND field \citep{RobertsGoss93, Vermot+25}.
The wavelengths at which the two transitions occur suffer from about the same degree of extinction, which means that their fluxes are readily comparable (e.g., \citealt{Fritz_2011}).
Thus, the weak residual peak near the CO$_2$ complex is very likely HI 16-10 emission.

\subsection{CO$_2$ Ice Abundances}
\begin{figure*}[htb]
    \centering
    \includegraphics[width=0.8\textwidth]{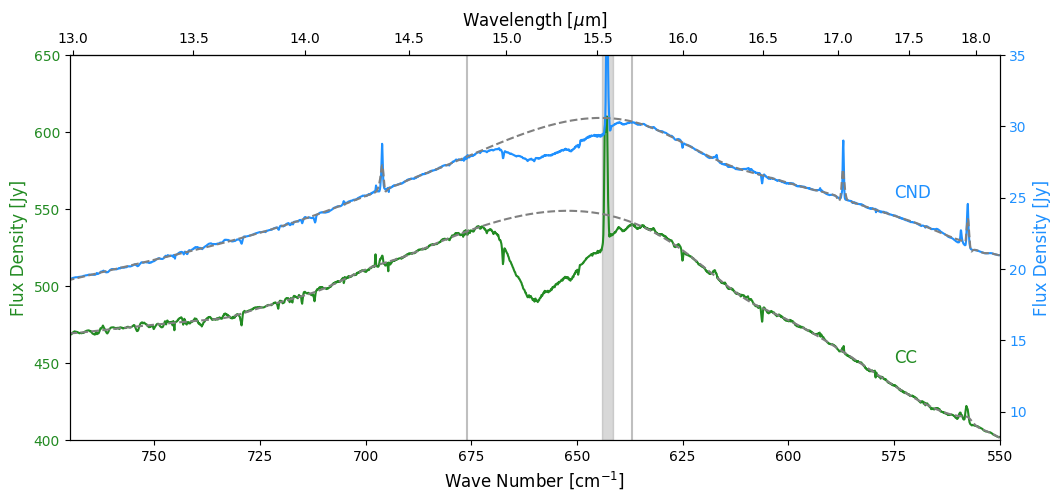}
    \caption{Integrated spectra of the two fields, centered around the CO$_2$ ice absorption feature. The gray dashed lines represent the derived baselines, the two narrow vertical lines represent the interval over which the equivalent widths are determined in order to infer the column densities, and the gray column represents the wavelengths over which the spectra are interpolated in order to prevent contamination by the Ne III emission line. Color coding is the same as in previous figures.}
    \label{fig:co2_ice_spectrum}
\end{figure*}

\begin{figure*}[htb]
    \centering
    \includegraphics[width=0.8\textwidth]{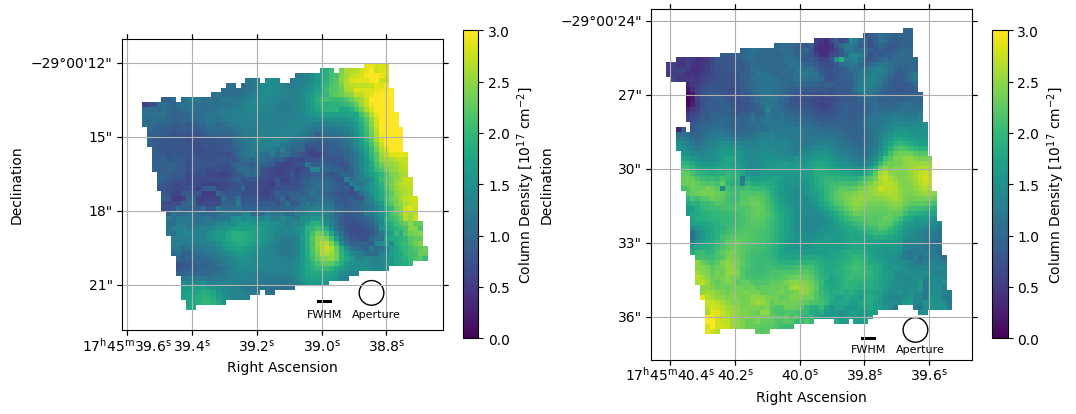}
    \caption{Column density maps of CO$_2$ ice in the two fields. Note that again the presence of CO$_2$ ice is ubiquitous in both fields, although there is significant spatial variation. The black circles on the bottom of the figures represent the aperture size chosen for the construction of these maps, and the adjacent bar represents the spatial resolution (FWHM) at 15 $\mu$m.}
    \label{fig:co2_ice_map}
\end{figure*}

We also observed a solid-state absorption feature caused by CO$_2$ ice in both of our fields.
This broad feature is shown in Figure~\ref{fig:co2_ice_spectrum}.
We chose the interval from 637 to 676 cm$^{-1}$ (14.79 to 15.70 $\mu$m) to estimate the column densities of CO$_2$ ice in the two fields, and adopting an integrated line strength of $1.1\times10^{-17}$cm molecule$^{-1}$ \citep{1995A&A...296..810G}, we estimated the column densities of CO$_2$ ice to be $(1.39\pm0.15)\times10^{17}$ cm$^{-2}$ in the CND field and $(1.84\pm0.20)\times10^{17}$ cm$^{-2}$ in the CC field.
This corresponds to an ice-to-gas abundance ratio of $\sim$90.
The CO$_2$ ice also has a significant spatial variation, which can be seen in Figure~\ref{fig:co2_ice_map}.
The maps were constructed in a manner similar to that used for the CO$_2$ gas, with each pixel assigned a column density value derived from the integrated spectrum within a circular aperture (radius of 0.5$''$, or 2.5 px) centered on it.


\section{Discussion} \label{sec:discussion}

Since the LSR radial velocities of CO$_2$ absorption from both fields are consistent with zero, and because the implied column densities of CO$_2$ between the two fields separated by $\sim20''$ are roughly similar, the most likely interpretation is that the observed absorption from both fields occurs in one or more foreground cloud(s).
Furthermore, since the molecular clouds near the Galactic Center have linewidths that are $\sim10$ times larger than those found in molecular clouds elsewhere in the Galaxy, and the 3 kpc arm has a radial velocity of $\sim-55$~km~s$^{-1}$, the observed absorptions likely arise from one or more molecular cloud(s) in a foreground spiral arm closer to the Sun than the 3 kpc arm.  
According to a recent study of Galactic extinction with GAIA \citep{gaia_extinction}, there appear to be candidate clouds at distances from the Sun of $\sim$0.15, $\sim$1.1, $\sim$2.0, $\sim$3.1, and $\sim$4.0 kpc.  Of course, more than one of these clouds could be contributing to the absorption features, and given the local velocity dispersion of molecular clouds in the Galactic disk on top of the near-zero radial velocity toward the Galactic center expected for Galactic rotation, we might expect a collection of multiple line-of-sight clouds to yield a relatively broad absorption line, such as what has been seen for radio emission from CO by \citet[][see their Figure~18]{GarciaP+16}.

Variations in the column density on the order of a factor of two between and within the two observed fields indicate that the cloud is somewhat clumpy.
The very roughly determined CO$_2$ rotational temperature in both fields is on the order of 30~K, somewhat less than the rotational temperature of C$_2$H$_2$ in both fields. 
Due to the solid-state nature of the CO$_2$ ice absorption, it is not possible to derive a radial velocity for the ice, but it is likely that it originates in the same cloud(s), and if so, the mean ice-to-gas ratio of 90 and some variation in the range of 60 to 140 are lower than but still consistent with previous simultaneous observations of the two components (\citealt{vanDishoeck+96}, \citealt{An_2009}).
Given that the aforementioned previous studies are on embedded young stars, it is not surprising that they see ice-to-gas ratio in the range of 10 to 100 while our observations have ratios that are $\gtrsim100$.


Our analysis shows a gas-phase column density ratio between C$_2$H$_2$ and CO$_2$ to be $\sim$ 1:3, and that between HCN and CO$_2$ to be roughly 3:2.
Interpretation of these ratios, which is beyond the scope of this paper, will require a treatment that accounts for the relative fraction of C$_2$H$_2$ and HCN that are in the solid phase.

This explanation applies to the absorption in the two fields, but a different explanation is required for the emission observed at the position of the aforementioned point-like emission in the column density map.
The location of this emission coincides with one of two known bright stellar sources dominating the continuum in MIRI/MRS channel 1 (R.A. $17^\text{h}45^\text{m}39\rlap{.}^{\text{s}}319$, Decl. $-29^{\circ}00'16\rlap{.}''302$ J2000). 
We identify these two point-like sources as the stars IRS~11 and IRS~11SW (\citealt{Zhu_2008}, \citealt{Ramirez_2000}, \citealt{2003yCat.2246....0C}): our channel 1 point sources are within $\sim0.3''$ and $\sim0.55''$, respectively, of the cataloged positions of these stars, and there are no other known sources in proximity.
Considering the astrometric errors of previous observations \citep{Zhu_2008} and the known astrometric precision of MIRI MRS (1-$\sigma$ deviation of $0.3''$ with cases of larger deviations\footnote{Astrometric calibration information from the JWST User Document, \url{https://jwst-docs.stsci.edu/jwst-calibration-status/miri-calibration-status/miri-mrs-calibration-status}.}), our measured positions of the stars are consistent with their cataloged positions.
The emission source is the only region in either field that shows CO$_2$ emission and is coincident with the point source that can be seen in 4.9 $\mu$m continuum in our data, identified as IRS~11SW.
Our measured radial velocity of the emission in the emission source ($-70\pm15$~km~s$^{-1}$) is close to the previously reported radial velocities for IRS 11SW ($-97\pm2$~km~s$^{-1}$, \citealt{Zhu_2008}).
Considering that the previously reported value has a systematic error of $\sim7$~km~s$^{-1}$, and with the uncertainties inherent in our method for deriving the emission profile, the velocity of the CO$_2$ is consistent with that of the star.
IRS~11SW is classified as an M3 III star by \citet{Zhu_2008}, a stellar type for which previous studies have reported CO$_2$ emission and, in some cases, absorption \citep{Ryde+99}, so the association of CO$_2$ emission with such a star is not unexpected.
We cross-matched our column density maps of the two fields with the positions of other known M- and K-type giants (there are 10 previously reported M- and K-type giants in the CC field), but there does not appear to be any appreciable feature, absorption or emission, at 14.98 $\mu$m at the locations of those stars.
What makes IRS 11SW different from other M-type giants in our fields of view when it comes to the presence of CO$_2$ features remains unclear, but an interesting possibility is raised by the fact that IRS 11SW is coincident with an X-ray source within 0.3 arcsec (\citealt{Muno_2003}, \citealt{Arendt+08}).
The star could therefore have the characteristics of a symbiotic binary \citep{Merc2025}, in which the CO$_2$ emission feature is generated in a wind from the M star or in an accretion flow onto a compact companion.

\section{Conclusions}  \label{sec:conclusions}
In this paper, we report the first observation of gas-phase CO$_2$ absorption and emission toward the bright background provided by the Galactic Center.
In addition to the observation of gas-phase CO$_2$, we also observed absorption features of gas-phase C$_2$H$_2$ and HCN, as well as CO$_2$ ice across the fields. 
We used a model assuming local thermodynamic equilibrium to fit the spectrum containing CO$_2$ $\nu_2$ transitions, and derive temperatures in the range of 20 to 50~K and densities of roughly 2$\times$10$^{15}$~cm$^{-2}$.
We find that HCN has  column densities similar to CO$_2$ but lower rotational temperatures ($\sim$10~K), while C$_2$H$_2$ has lower column densities (by a factor of $\sim$3) but higher temperatures ($\sim$100~K).
What these differences and similarities in temperature and column density represent about the line-of-sight distribution of temperature, density, and chemistry within the foreground cloud(s) will require higher spectral resolution observations and detailed modeling.
Similarly, further investigation targeting the ice counterparts is needed to establish where each absorption arises from the clouds and the corresponding distribution of grain temperature.
For all species, the measured column densities are consistent between our two exposures, whose central pointings are located 19.7$''$ apart.
 Combining these characteristics with the radial velocities consistent with zero, we conclude that the absorption from all three molecules arises from one or more extended foreground cloud(s).

Although most of the absorptions are attributable to foreground material, we also detect a CO$_2$ emission feature, coinciding with the location of IRS~11SW, a known M giant located in the Galactic Center possibly associated with an X-ray-emitting companion. 
M giant stars are known to be capable of producing circumstellar CO$_2$ emission. 
Moreover, we measured a high radial velocity of the emission at this location, which is consistent with the velocity reported in the literature for that star, strongly indicating an association. 

The observation of gas-phase CO$_2$ absorption at 15 $\mu$m is most favored to occur toward an extended, IR-bright region like the Galactic center.
Future observations of Galactic center environments and other IR-bright regions can enhance our understanding of how gas- and ice-phase CO$_2$ are distributed within foreground molecular clouds, and thereby improve our knowledge of gas-grain interactions within molecular clouds in general.
With a substantially larger number of CO$_2$ sightlines, it will be of interest to try to determine whether the gas/ice ratio is enhanced in the outermost low-extinction regions of clouds as a result of a greater amount of grain heating by the interstellar radiation field.

\begin{acknowledgments}
This work is based on observations made with the NASA/ESA/CSA James Webb Space Telescope.
The data were obtained from the Mikulski Archive for Space Telescopes at the Space Telescope Science Institute, which is operated by the Association of Universities for Research in Astronomy, Inc., under NASA contract NAS 5-03127 for JWST.
These observations are associated with program JWST GO 3166 (PI: A. Ciurlo).

Support for program JWST GO 3166 was provided by NASA through a grant from the Space Telescope Science Institute, which is operated by the Association of Universities for Research in Astronomy, Inc., under NASA contract NAS 5-03127.

AC, MRM and JQ acknowledge support from STScI grant JWST-GO-03166.001-A.
EACM gratefully acknowledges funding from the National Science Foundation under Award Nos. 2206509 and CAREER 2339670.

We also acknowledge the suggestions of the anonymous reviewer, who helped us improve the way we present our analysis.
\end{acknowledgments}

\appendix
\section{Uncertainty Calculations}
\subsection{Uncertainties on Temperature}
\label{appendix:A1}
\begin{figure}[htb]
    \centering
    \includegraphics[width=\linewidth]{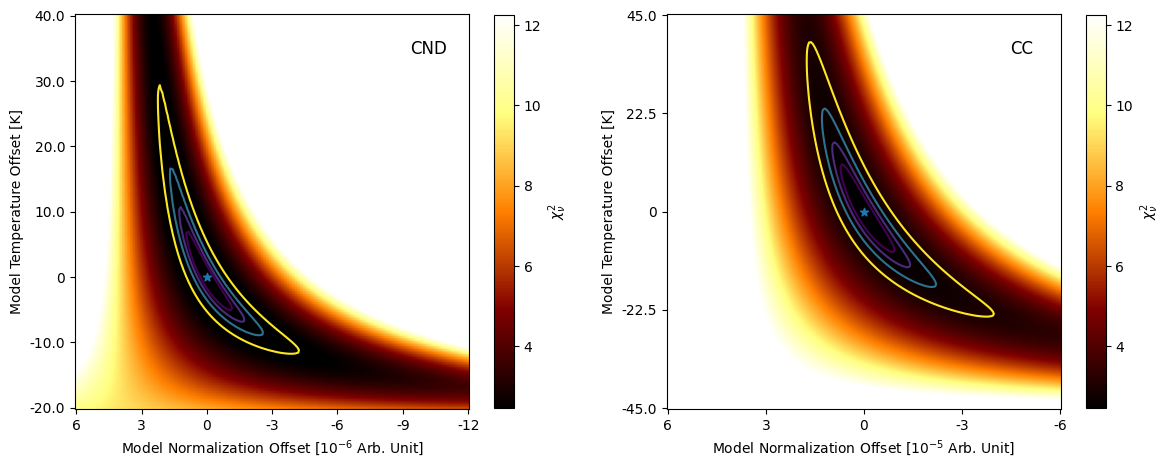}
    \caption{Reduced $\chi^2$ maps used to determine the uncertainties in temperature fits. The left map is for the CND pointing while the right is for the CC pointing. The stars denote the zero points, which are the best-fit values.}
    \label{fig:chi2_banana}
\end{figure}

The uncertainties in the temperature estimations are not straightforward to calculate, as the uncertainty in the covariant matrix of the model fit is purely statistical and is likely to grossly underestimate the total temperature uncertainty.
To derive a more faithful uncertainty, we calculated the reduced $\chi^2$ for a grid of parameters for both spectra for each of the two pointings.
The intrinsic line width and the radial velocity were kept the same when estimating the $\chi^2_\nu$ map.
Following the approach in \citet{Boonman2003}, we traced the contours of $\chi^2_\nu$ corresponding to a $2.5\%, 5\%, 10\%,\text{ and }22.5\%$ increase over the minimal $\chi^2_\nu$, shown in Figure~\ref{fig:chi2_banana}.
The extrema of the contour (outer yellow line in the figure) corresponding to a $22.5\%$ increase are taken to be the uncertainty in the modeled temperature.
We used the extrema instead of the y-intercept of a vertical line crossing the minimal $\chi^2_\nu$ because the temperature and normalization are correlated, i.e. a more negative (stronger) normalization could compensate for the lowering of temperature.

\subsection{Uncertainties on Column Density}
\label{appendix:A2}
The two quantities that affect the calculation of the uncertainty on the column density of CO$_2$ are the uncertainties in temperature and in equivalent width.
The former affects the contribution of different $J$ levels to the total column density, and the latter determines general normalization.
To assess the influence of the temperature uncertainty on the column density, we used the upper and lower bounds of the uncertainty on temperature to calculate two new column densities to see how different they are from the original result.
But for both spectra from the whole fields of view, the uncertainty in column density derived from the uncertainty in temperature is less than 0.1$\%$, overwhelmed by the uncertainty that comes from the calculation of the equivalent width by a factor of 50, so this uncertainty is not taken into account.

The uncertainty in equivalent width has two main components: the ``statistical'' uncertainty coming from the noise in the data and the uncertainty that comes from different choices of the baseline, which are somewhat arbitrary. 
We calculated the statistical contribution to the uncertainty using the standard deviation derived from the baseline-subtracted, absorption-free regions of the spectra and determined it to be $\sqrt{N}\sigma$, where N is the number of data points in the calculation and $\sigma$ is the standard deviation.
We estimated the contribution of the baseline choice as $4\sqrt{N}\sigma$ where the standard deviation is derived from the baseline-subtracted, absorption-free regions of the spectra after trying different baseline choices.
This means that the overall uncertainty in the column density is $\sim4\sqrt{N}\sigma$, with the choice of baseline dominating the uncertainty.
Considering all these factors, we found a $16\%$ uncertainty in the column density reported for the CND pointing, and a $12\%$ uncertainty in that reported for the CC pointing.

In addition to what is reported in the main body of this work, the column density of CO$_2$ can also be calculated using the LTE model alone.
One approach is to use the R(4) transition and assume thermodynamic equilibrium to calculate the total column density.
Using only the LTE model fit parameters and taking instrumental broadening effect into consideration, we calculated the optical depth and intrinsic line width of that given transition.
Then we calculated the column density of the gas in the $J=4, \nu=0$ lower energy state along with the total column density.
This method yields a difference of $<1\%$ for the CND pointing and $\sim5\%$ for the CC pointing compared to the values reported in Table \ref{table:params}.

The result of this approach agrees with the value derived using equivalent width only, so it serves as a confirmation.
The column densities derived using equivalent width are presented in the main body, as those values rely less on the model and are not sensitive to the intrinsic line width, unlike the two methods presented here, which use the poorly constrained intrinsic line width to derive the optical depth.

\subsection{Uncertainties on Radial Velocity and Intrinsic Velocity Dispersion}
To estimate the uncertainty in the fitted Doppler velocity, the strongest P- and R- branch lines are fitted individually using a Gaussian profile.
The standard deviations of the Doppler velocities derived using single lines are taken to be the ``statistical'' uncertainty in the radial velocity and correspond to 4~km~s$^{-1}$ for the CND, and 5~km~s$^{-1}$ for the CC.
In addition, instrumental and calibration uncertainties are also taken into account, which correspond to 7~km~s$^{-1}$ of systematic uncertainty \citep{Argyriou2023}.

\section{CO$_2$ Temperature and Radial Velocity Spatial Variations}
\label{appendix:B}
One essential attribute of an integral field unit such as MIRI/MRS is spatial resolution.
However, because of the low signal-to-noise ratio of CO$_2$ features in small apertures, the parameter maps derived with the LTE model (radial velocity and temperature) suffer from large uncertainties. 
Moreover, temperature and radial velocity are correlated: the temperature determines the shape of the absorption, and thus affects the best-fit radial velocity. 
Because of this correlation, the maps showing their distribution across the fields of view constructed with freely varying parameters do not show variations significantly larger than the statistical uncertainties.
It is worth noting that if we assume a uniform temperature (as derived using the integrated spectrum, a not unreasonable assumption if the cloud is indeed foreground), the radial velocity maps constructed are fairly smooth with small gradients, again comparable to the statistical uncertainty.

\section{Spatial coincidence of CO$_2$ emission and IRS~11SW}
\begin{figure}[htb]
    \centering
    \includegraphics[width=\textwidth]{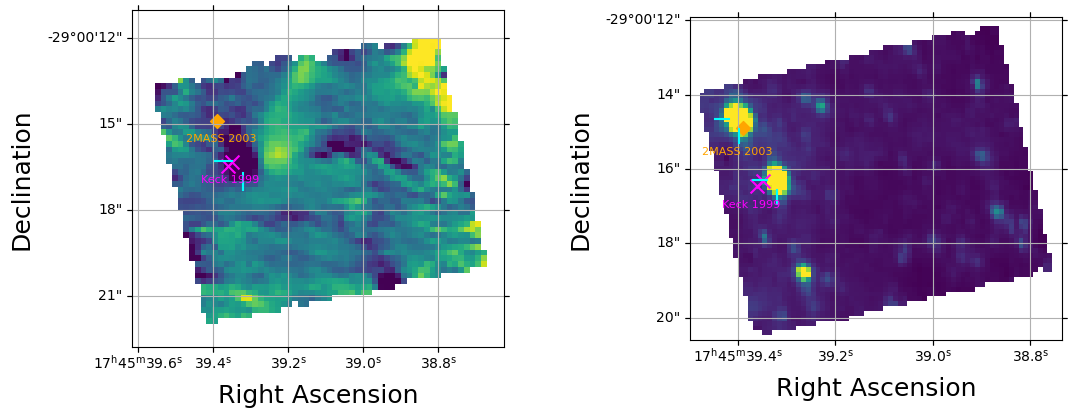}
    \caption{Column density map (left) compared to the continuum map (right), where the two bright sources are identified as IRS 11 and IRS 11SW. Previously cataloged positions are marked by the orange diamond and magenta crosses, respectively, and the cyan bars mark out the centroids, found using \texttt{centroid\_1dg}. The cataloged positions of IRS~11 and IRS~11SW come from the 2003 2MASS survey \citep{2003yCat.2246....0C} and the W.M. Keck Observatory in 1999 (\citealt{Zhu_2008}, \citealt{Figer_2003}).}
    \label{fig:dip_and_emission}
\end{figure}

\begin{figure}[htb]
    \centering
    \includegraphics[width=\textwidth]{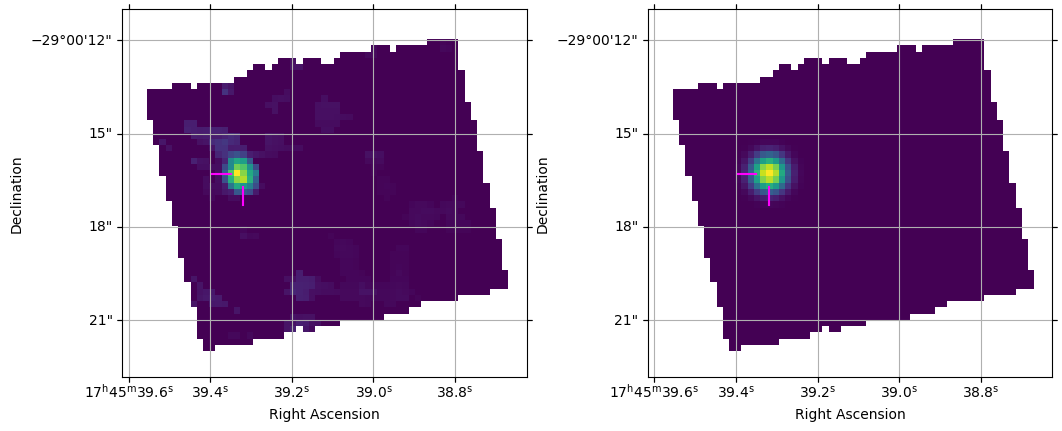}
    \caption{CO$_2$ emission strength (left) versus PSF (FWHM$\sim$0.6$''$) treated with the circular aperture method. The emission component of CO$_2$ seems to be produced by a point source.}
    \label{fig:emission_and_FWHM}
\end{figure}

For completeness, here we present the graphical evidence in Figure~\ref{fig:dip_and_emission} for the identification of the association between the ``dip'' and the star IRS~11SW.
In the right panel is the continuum of channel 1 (toward the shortest wavelength, $\sim4.9\mu$m), which is dominated by stellar sources.
The two brightest sources are close to the previously reported positions of stars IRS~11 and IRS~11SW, shown by the orange diamond and magenta crosses, respectively. We also found the centroid of the two sources, shown by the cyan crosses.
The same symbols are overlaid on the CO$_2$ column density map in the left panel, and we see that IRS~11SW coincides with the position of the ``dip'', while IRS~11 does not associate with any outstanding feature, even though it is also an M-type giant.
The emission is consistent with a point source, as shown in Figure~\ref{fig:emission_and_FWHM}, further supporting the conjecture that it is produced by the star.

\section{NLTE model for $\mathrm{CO}_2$}\label{NLTE_Model}

In the context of accretion disks and planetary formation, \citet{2017A&A...601A..36B} have shown that non-LTE effects (NLTE) might affect $\mathrm{CO}_2$ excitation and that radiative pumping in the infrared followed by radiative cascades might not be negligible.
To test whether this should be taken into account in the context of our observations, we developed an NLTE model of $\mathrm{CO}_2$ derived from the model presented in \citet{2025A&A...699A.217L} for $\mathrm{C}_2$.
The model includes excitation and de-excitation by collisions, spontaneous radiative decay, radiative pumping to high energy levels followed by cascades and possibly excitation at chemical formation.
All relevant equations and details are presented in \citet{2025A&A...699A.217L}.

In the case of $\mathrm{CO}_2$, computations are limited to the first $21$ rotational levels (up to $J=40$), just below the first excited vibrational level.
Collision rate coefficients with $\mathrm{He}$ come from \citet{2022JChPh.156j4303G}, collisions with $\mathrm{H}$ and $\mathrm{H}_2$ have been scaled from them using reduced mass.
The quadrupolar transition probabilities come from \citet{2021JChPh.154u1104Y}.
All other radiative transitions are taken from the HITRAN database, including $172\,376$ transitions between $19\,043$ levels.
These transitions allow us to compute cascade coefficients from all reachable levels above $J=40$ down to one of the computed levels by any possible path (see Appendix B of \citealt{2025A&A...699A.217L} for details).
Possible excitation at formation is accounted for, although this should have a negligible impact here.

\begin{figure}[htb]
    \centering
    \includegraphics[width=0.7\columnwidth]{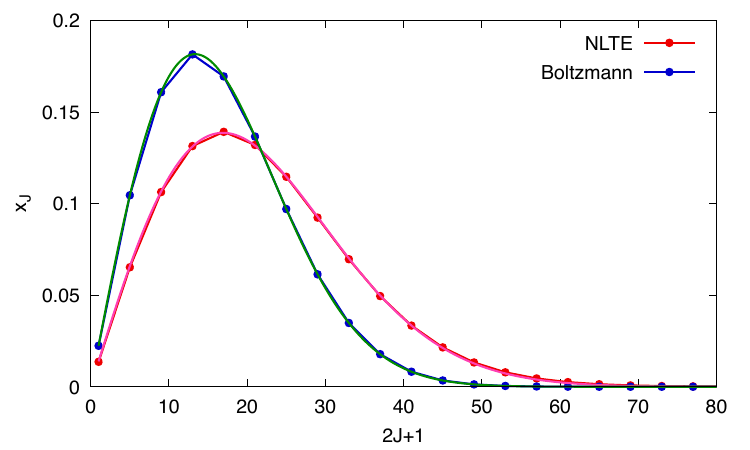}
    \caption{Comparison of LTE and NLTE populations for $n_\mathrm{H} = 30\,\mathrm{cm}^{-3}$, $T = 50\,\mathrm{K}$ and $G_0 = 100$. This configuration has a ``distance'' of 0.051 to LTE.}
    \label{fig:NLTE_Pop}
\end{figure}

\begin{figure}[htb]
    \centering
    \includegraphics[width=0.8\columnwidth]{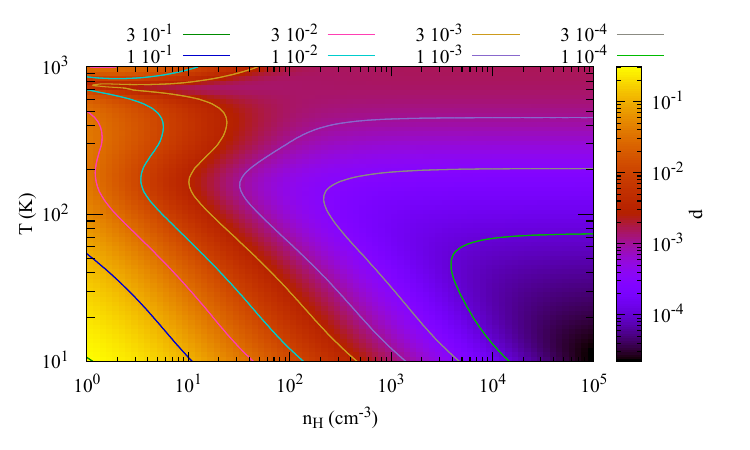}
    \caption{``Distance'' (as defined in Equation \ref{eq:dist}) to LTE for $G_0 = 10$.}
    \label{fig:Grid_NLTE}
\end{figure}

Figure~\ref{fig:NLTE_Pop} shows a comparison of populations in an extreme case for an ISRF scaled up by a factor of $100$ and diffuse cloud conditions.
This is rather unrealistic for the foreground clouds that we deem responsible for the observed absorption, but allows us to visualize the possible NLTE impact of radiative cascades.
We see that both populations can be very well approximated by an analytical fit:

\begin{equation}\label{eq:fit}
    f(x) = a\, x \, \exp{\left( -b x^c \right)},
\end{equation}

where $x = 2J+1$.
This equation is exact for a Boltzmann population (although not in its standard form) and follows closely the NLTE results.
It allows us to define a ``distance'' $d$ from LTE by:

\begin{equation}\label{eq:dist}
    d^2 = (a_1 - a_2)^2 + \left( \frac{b_1 - b_2}{0.15} \right)^2 + \left( \frac{c_1 - c_2}{2} \right)^2,
\end{equation}

where index $1$ refers to LTE and index $2$ to NLTE.
The numerical factors are adjusted so that all $3$ parameters are of the order of $1$.
We kept $c_1 = 2$ for all models.
Figure~\ref{fig:Grid_NLTE} shows the resulting ``distance'' for $G_0 = 10$.
We find that NLTE effects become less pronounced with increasing density and decreasing radiation field.
Given that the observed absorbing medium is likely in the near foreground, we chose relatively conservative parameters (which likely yield more NLTE effect than the actual ones) typical of a diffuse medium where $T=50$ K, $n_\text{H}=100 \text{ cm}^{-3}$, and $G_0=3$.
Using these parameters, we calculated the ``distance'' in this situation to be $\sim4.3\times10^{-4}$, and concluded that NLTE effects can be safely neglected in the context of the present study.

\bibliography{bibliography}{}
\bibliographystyle{aasjournal}

\end{document}